\RequirePackage{fix-cm}
\documentclass[smallextended]{svjour3} % onecolumn (second format)
\smartqed % flush right qed marks, e.g. at end of proof
\usepackage{graphicx}
\usepackage{amsmath}
\usepackage{mathcomp}

 \journalname{Exp Astron}
 
\usepackage{xcolor}

\usepackage[colorlinks=true,breaklinks=true,linkcolor=magenta,citecolor=blue,urlcolor=green]{hyperref}

\newcommand\ctscm{counts~cm$^{-2}$} 
\newcommand\ergs{erg~s$^{-1}$}

\newcommand\ergcm{erg~cm$^{-2}$}

\newcommand\apj{{Astrophys.\ J.}}%
\newcommand\apjl{{Astrophys.\ J.}}%
\newcommand\apjs{{ApJS}}%
\newcommand\aap{{Astron. Astrophys.}}%
\newcommand\mnras{{Mon.\ Not.\ R.\ Astron.\ Soc.}}%
\newcommand\nat{{Nature}}%
\newcommand\grl{{Geophys.~Res.~Lett.}}%
\newcommand\cjaa{{Chinese J. Astron. Astrophys.}}%
\newcommand\prl{{Phys.~Rev.~Lett.}}%
 \newcommand\procspie{{Proc.~SPIE}}%
 
\begin{document}

\title{GRID: a Student Project to Monitor the Transient Gamma-Ray Sky in the Multi-Messenger Astronomy Era}

\author{
Jiaxing Wen$^{1,2}$ \and
Xiangyun Long$^{3}$ \and
Xutao Zheng$^{1}$ \and
% alphabetically
Yu An$^{4}$ \and
Zhengyang Cai$^{3}$ \and
Jirong Cang$^{1}$ \and
Yuepeng Che$^{5}$ \and
Changyu Chen$^{1}$ \and
Liangjun Chen$^{4}$ \and
Qianjun Chen$^{1}$
% ...
\and
$^{1}$Department of Engineering Physics and Center for Astrophysics, Tsinghua University, Beijing 100084, China \and
$^{2}$Science and Technology on Plasma Physics Laboratory, Laser Fusion Research Center, Chinese Academy of Engineering Physics, Mianyang 621900, China \and
$^{3}$Department of Physics and Center for Astrophysics, Tsinghua University, Beijing 100084, China \and
$^{4}$Guangxi Key Laboratory for Relativistic Astrophysics, Department of Physics, Guangxi University, Nanning, Guangxi 530004, China \and
$^{5}$School of Electronic and Information Engineering, Ningbo University of Technology, Ningbo, Zhejiang 315211, China \and
$^{6}$Department of Instruments Science \& Engineering, Shanghai Jiaotong University, Shanghai 200240 \and
$^{7}$School of Materials Science \& Chemical Engineering, Ningbo University, Ningbo 315211, China \and
$^{8}$School of Astronomy and Space Science, Nanjing University, Nanjing, Jiangsu 210093, China \and
$^{9}$School of Aerospace Engineering, Tsinghua University, Beijing 100084, China \and
$^{10}$College of Physical Science and Technology, Sichuan University, Chengdu 610065, China \and
$^{11}$School of Physics and Astronomy, Sun Yat-Sen University, Zhuhai 519082, China \and
$^{12}$Department of Astronomy, Xiamen University, Xiamen, Fujian 361005, China \and
$^{13}$School of Mathematics and Physics, University of Science and Technology Beijing, Beijing 100083, China \and
$^{14}$Department of Foreign Languages and Literatures, Tsinghua University, Beijing 100084, China \and
$^{15}$Department of Electrical Engineering, Tsinghua University, Beijing 100084, China \and
$^{16}$School of Software, Tsinghua University, Beijing 100084, China \and
$^{17}$Faculty of Electrical Engineering and Computer Science, Ningbo University, Ningbo 315211, China \and
$^{18}$Department of Modern Physics, School of Physical Sciences, University of Science and Technology of China, Hefei, Anhui 230026, China \and
$^{19}$Department of Astronomy, Beijing Normal University, Beijing 100875, China \and
$^{20}$Department of Astronomy, School of Physical Sciences, University of Science and Technology of China, Hefei, Anhui 230026, China \and
$^{21}$Purple Mountain Observatory, Chinese Academy of Sciences, Nanjing 210008, China \and
$^{22}$Department of Astronomy, Tsinghua University, Beijing 100084, China \and
$^{23}$School of Physical Science and Technology, Lanzhou University, Lanzhou 730000, China \and
$^{24}$Kavli Institute for Astronomy and Astrophysics and Department of Astronomy, School of Physics, Peking University, Beijing 100087, China \and
$^{25}$School of Physics and Technology, Wuhan University, Wuhan 430072, China \and
$^{26}$Key Laboratory for Particle Astrophysics, Institute of High Energy Physics, Chinese Academy of Sciences, Beijing 100049, China \and
$^{27}$Institute of Theoretical Physics, Chinese Academy of Sciences, Beijing 100190, China \and
$^{28}$Department of Physics, The University of Hongkong, Hong Kong, China \and
$^{29}$National Astronomical Observatories, Chinese Academy of Sciences, Beijing 100012, China \and
$^{30}$State Key Laboratory of Lunar and Planetary Sciences, Macau University of Science and Technology, Macau, China \and
$^{31}$Shanghai Astronomical Observatory, Chinese Academy of Sciences, Shanghai 200030, China \and
$^{32}$Institute of Astrophysics, Central China Normal University, Wuhan 430079, China \and
$^{33}$Key Laboratory of Modern Astronomy and Astrophysics (Nanjing University), Ministry of Education, China \and
$^{34}$Department of Physics and Astronomy, University of Nevada, Las Vegas, NV 89154, USA}

\institute{
H.~Feng \at \email{hfeng@tsinghua.edu.cn} \and
M.~Zeng \at \email{mingzeng@tsinghua.edu.cn} \and
B.-B.~Zhang \at \email{bbzhang@nju.edu.cn} \and
$^{1}$Department of Engineering Physics and Center for Astrophysics, Tsinghua University, Beijing 100084, China \and
$^{2}$Science and Technology on Plasma Physics Laboratory, Laser Fusion Research Center, Chinese Academy of Engineering Physics, Mianyang 621900, China \and
$^{3}$Department of Physics and Center for Astrophysics, Tsinghua University, Beijing 100084, China \and
$^{4}$Guangxi Key Laboratory for Relativistic Astrophysics, Department of Physics, Guangxi University, Nanning, Guangxi 530004, China \and
$^{5}$School of Electronic and Information Engineering, Ningbo University of Technology, Ningbo, Zhejiang 315211, China \and
$^{6}$Department of Instruments Science \& Engineering, Shanghai Jiaotong University, Shanghai 200240 \and
$^{7}$School of Materials Science \& Chemical Engineering, Ningbo University, Ningbo 315211, China \and
$^{8}$School of Astronomy and Space Science, Nanjing University, Nanjing, Jiangsu 210093, China \and
$^{9}$School of Aerospace Engineering, Tsinghua University, Beijing 100084, China \and
$^{10}$College of Physical Science and Technology, Sichuan University, Chengdu 610065, China \and
$^{11}$School of Physics and Astronomy, Sun Yat-Sen University, Zhuhai 519082, China \and
$^{12}$Department of Astronomy, Xiamen University, Xiamen, Fujian 361005, China \and
$^{13}$School of Mathematics and Physics, University of Science and Technology Beijing, Beijing 100083, China \and
$^{14}$Department of Foreign Languages and Literatures, Tsinghua University, Beijing 100084, China \and
$^{15}$Department of Electrical Engineering, Tsinghua University, Beijing 100084, China \and
$^{16}$School of Software, Tsinghua University, Beijing 100084, China \and
$^{17}$Faculty of Electrical Engineering and Computer Science, Ningbo University, Ningbo 315211, China \and
$^{18}$Department of Modern Physics, School of Physical Sciences, University of Science and Technology of China, Hefei, Anhui 230026, China \and
$^{19}$Department of Astronomy, Beijing Normal University, Beijing 100875, China \and
$^{20}$Department of Astronomy, School of Physical Sciences, University of Science and Technology of China, Hefei, Anhui 230026, China \and
$^{21}$Purple Mountain Observatory, Chinese Academy of Sciences, Nanjing 210008, China \and
$^{22}$School of Physical Science and Technology, Lanzhou University, Lanzhou 730000, China \and
$^{23}$Kavli Institute for Astronomy and Astrophysics and Department of Astronomy, School of Physics, Peking University, Beijing 100087, China \and
$^{24}$School of Physics and Technology, Wuhan University, Wuhan 430072, China \and
$^{25}$Key Laboratory for Particle Astrophysics, Institute of High Energy Physics, Chinese Academy of Sciences, Beijing 100049, China \and
$^{26}$Institute of Theoretical Physics, Chinese Academy of Sciences, Beijing 100190, China \and
$^{27}$Department of Physics, The University of Hongkong, Hong Kong, China \and
$^{28}$National Astronomical Observatories, Chinese Academy of Sciences, Beijing 100012, China \and
$^{29}$State Key Laboratory of Lunar and Planetary Sciences, Macau University of Science and Technology, Macau, China \and
$^{30}$Shanghai Astronomical Observatory, Chinese Academy of Sciences, Shanghai 200030, China \and
$^{31}$Institute of Astrophysics, Central China Normal University, Wuhan 430079, China \and
$^{32}$Key Laboratory of Modern Astronomy and Astrophysics (Nanjing University), Ministry of Education, China \and
$^{33}$Department of Physics and Astronomy, University of Nevada, Las Vegas, NV 89154, USA
}

\authorrunning{Wen et al.} % if too long for running head
\titlerunning{GRID: A Student Project}

\date{Received: XX XXX 2019 / Accepted: XX XXX 2019}

\maketitle

\begin{abstract}

The Gamma-Ray Integrated Detectors (GRID) is a space mission concept dedicated to monitoring the transient gamma-ray sky in the energy range from 10 keV to 2 MeV using scintillation detectors onboard CubeSats in low Earth orbits. The primary targets of GRID are the gamma-ray bursts (GRBs) in the local universe. The scientific goal of GRID is, in synergy with ground-based gravitational wave (GW) detectors such as LIGO and VIRGO, to accumulate a sample of GRBs associated with the merger of two compact stars and study jets and related physics of those objects. It also involves observing and studying other gamma-ray transients such as long GRBs, soft gamma-ray repeaters, terrestrial gamma-ray flashes, and solar flares. With multiple CubeSats in various orbits, GRID is unaffected by the Earth occultation and serves as a full-time and all-sky monitor. Assuming a horizon of 200~Mpc for ground-based GW detectors, we expect to see a few associated GW-GRB events per year. With about 10 CubeSats in operation, GRID is capable of localizing a faint GRB like 170817A with a 90\% error radius of about 10 degrees, through triangulation and flux modulation. GRID is proposed and developed by students, with considerable contribution from undergraduate students, and will remain operated as a student project in the future. The current GRID collaboration involves more than 20 institutes and keeps growing. On August 29th, the first GRID detector onboard a CubeSat was launched into a Sun-synchronous orbit and is currently under test.  


\end{abstract}

\section{Introduction}
\label{sec:intro}

The historical gravitational wave event, GW170817, detected by the Laser Interferometer Gravitational-Wave Observatory (LIGO) and its associated electromagnetic (EM) counterparts mark a new era in astronomy, the so-called multi-messenger astronomy with flying colors~\cite{Abbott2017,Abbott2017a}. 
The joint detection of GW170817 and GRB 170817A has been long-sought for decades to confirm the progenitor models of short-duration GRBs that invoke mergers of two compact stellar objects including the most favorable neutron star-neutron star (NS-NS) and neutron star-black hole (NS-BH) systems ~\cite{Paczynski1986,Eichler1989,Zhang2009,Abbott2017b,Ajello2018,Savchenko2017,Zhang2018,Goldstein2017,Pozanenko2018}. Although a few candidates (e.g, GRB 130603B~\cite{Tanvir2013} and GRB 050709~\cite{Jin2016}) had been previously proposed, the optical and near-infrared counterpart that was revealed in unprecedented details a few hours after GRB 170817A was indubitably recognized as the first confirmed kilonova in history and was remarkably consistent with the theoretical expectation~\cite{Li1998,Metzger2010,Arcavi2017,Smartt2017}. Such observations also suggest the kilonova be the main channel to produce heavy elements in the universe via the r-process~\cite{Pian2017,Drout2017}. Future identifications of similar GW-GRB events will offer us a promising means to understanding the physics of short GRB as well as measuring the Hubble constant~\cite{Abbott2017c}, constraining the equation of state of neutron stars and other fundamental physics such as the violation of Lorentz invariance~\cite{Abbott2017}.

The gamma-ray emission of GRB 170817A is puzzling. If the distance were unknown, it would appear to be an otherwise normal short burst with slightly weak flux and fluence \cite{Zhang2018}. However, with its known distance of $40^{+8}_{-14}$~Mpc, its peak luminosity of $\sim 1.6\times10^{47}$~\ergs\ marks it the least luminous short GRB known so far, which is about 3 orders of magnitude lower than those of typical GRBs detected before. For many other GRBs without a redshift measurement, it is unclear if there are similar low-luminosity ones like GRB 170817A \cite{Yue2018}, or if the GRB luminosity function has a smooth distribution all the way from $10^{50}$~\ergs\ down to $10^{47}$~\ergs. The jet physics on how to produce the $\gamma$-ray emission is also controversial. It is suggested that the burst was due to a jet breaking out of a cocoon-like shell~\cite{Kasliwal2017,Gottlieb2018} or a structured jet viewed at a large off-axis angle \cite{Zhang2018}. More events are needed to understand the physics about the jet formation and structure and the final product after the merger of neutron stars \cite{Ai2018,Beniamini2019}. 

Thanks to the small distance of GW170817, within which the total number of galaxies in available catalogs is limited, optical telescopes can target at known galaxies rather than execute a blind search in the $\sim$28 square degrees of the sky \cite{Abbott2017,Abbott2017a}. This is perhaps one of the most favorable factors that lead to the successful identification of the optical counterpart but is not guaranteed for future GW events. Future upgrades will further increase the sensitivity and horizon for the GW detectors. The advanced LIGO is expected to have a horizon of $\sim$200~Mpc for binary neutron star merger events upon 2020~\cite{Abbott2016}. In that case, the search of the EM counterparts may become challenging, and a better accuracy in localization will help.

As a distributed system, the GRID project aims to build a full-time and all-sky gamma-ray detection network in low Earth orbits, without Earth occultation or interruptions due to South Atlantic Anomaly (SAA) passes. We note that quite a few mission concepts similar to GRID have been proposed and are under development \cite{Racusin2017,Werner2018,Fuschino2018,Chattopadhyay2018,Hui2018}.  Compared with GRID, BurstCube \cite{Racusin2017} and CAMELOT \cite{Werner2018} adopt a similar detection technology; HERMES is greatly enhanced in the number of satellites with an extension to the soft X-ray band using silicon detectors \cite{Fuschino2018}; BlackCAT \cite{Chattopadhyay2018} is designed to have a coded mask for localization; MoonBEAM \cite{Hui2018} will be deployed into a cislunar orbit with a much longer baseline. 

Besides the primary goal of detecting GRBs, the GRID network is also sensitive to other high energy transients. For example, it can observe the soft $\gamma$-ray repeaters (SGRs), which are thought to be connected with the magnetic activity of extremely magnetized neutron stars (a.k.a, magnetars). GRID is capable of monitoring SGRs from known magnetars and finding new sources. Another example is the Terrestrial $\gamma$-ray flashes (TGFs), which are millisecond $\gamma$-ray flashes originated from the Earth atmosphere, likely linked to thunderstorm activities \cite{Cummer2011}. The same process that produces TGFs may also produce terrestrial electron beams (TEBs)~\cite{Connaughton2013}. GRID is expected to detect a large sample of TGFs. It may also detect TEBs captured by the Earth magnetosphere by correlating triggers from CubeSats at different locations. Moreover, GRID is capable of monitoring solar activities in the $\gamma$-ray band \cite{Ackermann2012}.

Thanks to the technical readiness of $\gamma$-ray detectors, GRID is suitable for a student project. GRID was initially proposed and developed by students, with a considerable contribution from undergraduate students. The current student leader of GRID, Jiaxing Wen, was first a senior undergraduate student when GRID was proposed and now is a graduate student in his first year. The development work, including the scientific justification, instrument design, detector assembly and tests, laboratory calibrations, and space qualification experiments, were all led and accomplished by the student team. The purpose of GRID is twofold. Besides its scientific goals, we hope to attract excellent students from different disciplines into astrophysics and train them on how to organize and participate in a multi-discipline collaboration. The students can learn how to build a real science project that covers hardware, data and science. Ultimately, GRID is a scientific collaboration that accepts students and scientists from all over the world; the members can launch their own detectors, share the data, and produce the science results under certain agreements. 

\begin{figure}
\includegraphics[width=0.6\textwidth]{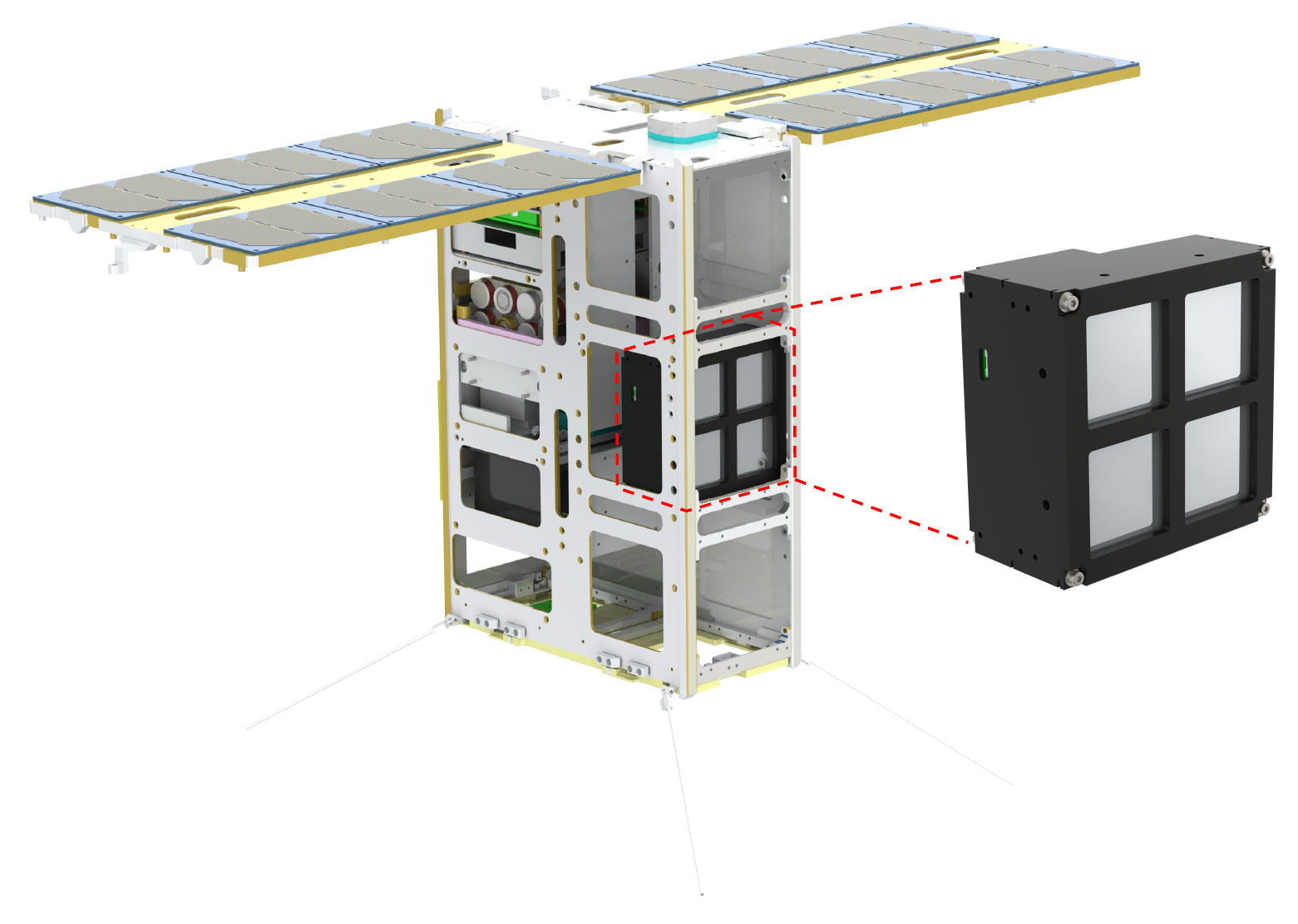}
\caption{Schematic drawing of the 6U CubeSat developed by Spacety with the first GRID detector on it.}
\label{fig:cubesat}
\end{figure}

On October 29th, 2018, the first prototype of the GRID detector onboard a CubeSat (Figure~\ref{fig:cubesat}) was launched into space in a Sun-synchronous orbit by a commercial satellite company. The detector is under test and the results will be reported elsewhere. In this paper, we will introduce the GRID concept, including the scientific objectives, instrument design, laboratory test results, detector performance, and the framework of the collaboration. 

\section{The GRID concept}

The primary scientific goal of GRID is to detect the gamma-ray events associated with future GW events that can be detected by ground-based facilities such as LIGO and Virgo. We plan to deploy at least ten identical $\gamma$-ray detectors in low Earth orbits (500-600 km) using CubeSats. Compared with a single, large satellite, GRID is unaffected by Earth occultation and can cover the whole sky. On each CubeSat, a scintillation detector with an effective energy range from 10 keV to 2 MeV is used to monitor GRBs and other $\gamma$-ray transients. A single detector has no localization capabilities, while a $\gamma$-ray source can be localized if it is jointly detected by at least three CubeSats. The position can be reconstructed by means of triangulation, similar to the interplanetary network (IPN)\footnote{http://www.ssl.berkeley.edu/ipn3/}~\cite{Hurley2016}, or by differences in flux observed by different detectors (hereafter referred to as ``flux modulation''), similar to the methods used by the Gamma-ray Burst Monitor (GBM) onboard the Fermi Satellite \cite{Connaughton2015} and the Burst And Transient Source Experiment (BATSE) onboard the Compton Gamma Ray Observatory~(CGRO). In order to minimize the instrumental effect, identical detectors will be used for the GRID network. The localization uncertainty may be subject to additional systematic errors, e.g., due to inaccurate mass modeling of the CubeSat structure.

In the following, we will describe the current design of the system. With two or three in-orbit tests, we aim to deliver a standard hardware/software design that is ready to be duplicated and launched by the collaboration members, which will help expand the GRID network and achieve the scientific outcome.

\subsection{Detector}

\begin{figure}
\includegraphics[width=0.5\textwidth]{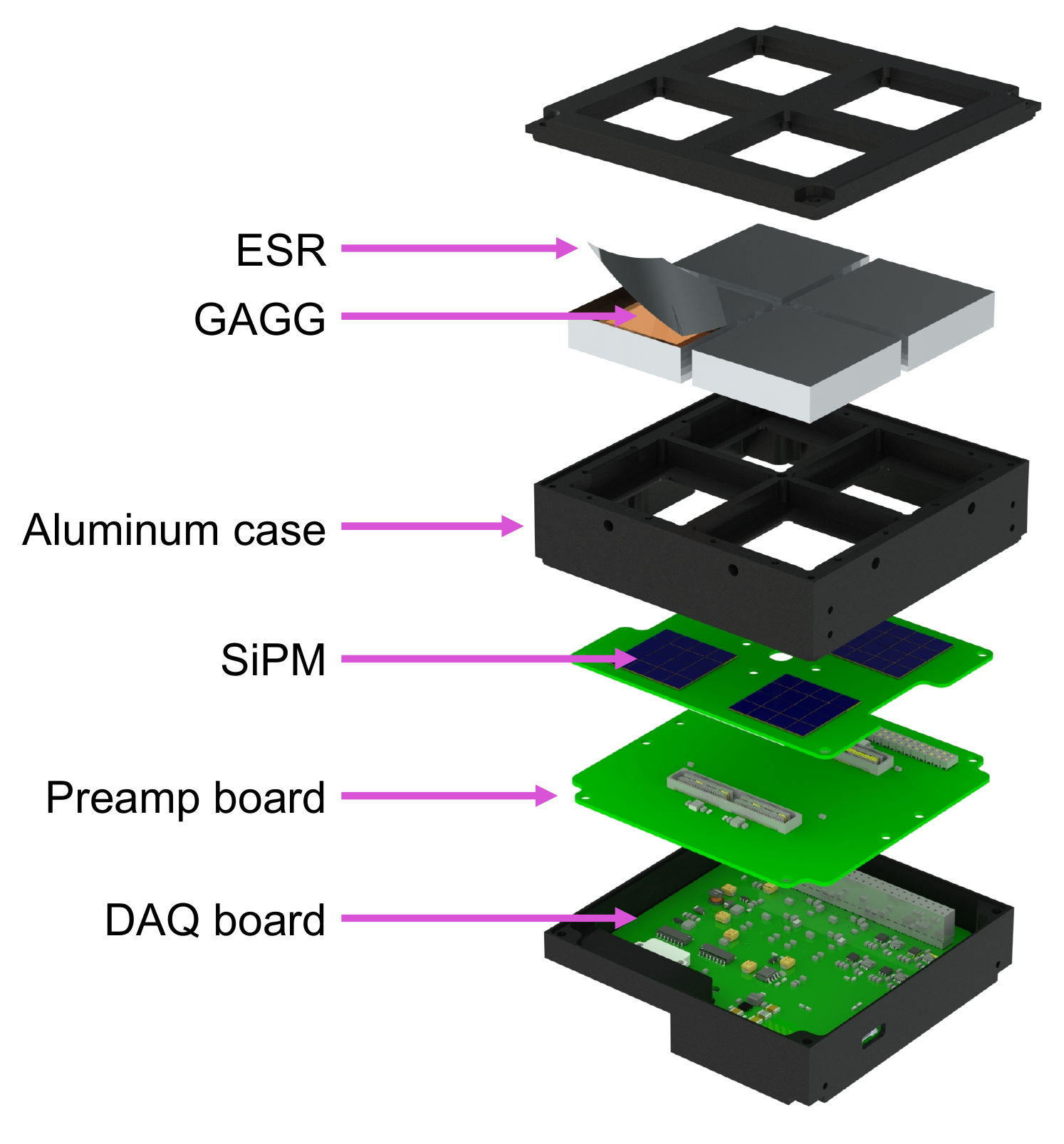}
\caption{Structure of the GRID detector. }
\label{fig:detector}
\end{figure}

\begin{figure}
\includegraphics[width=0.6\textwidth]{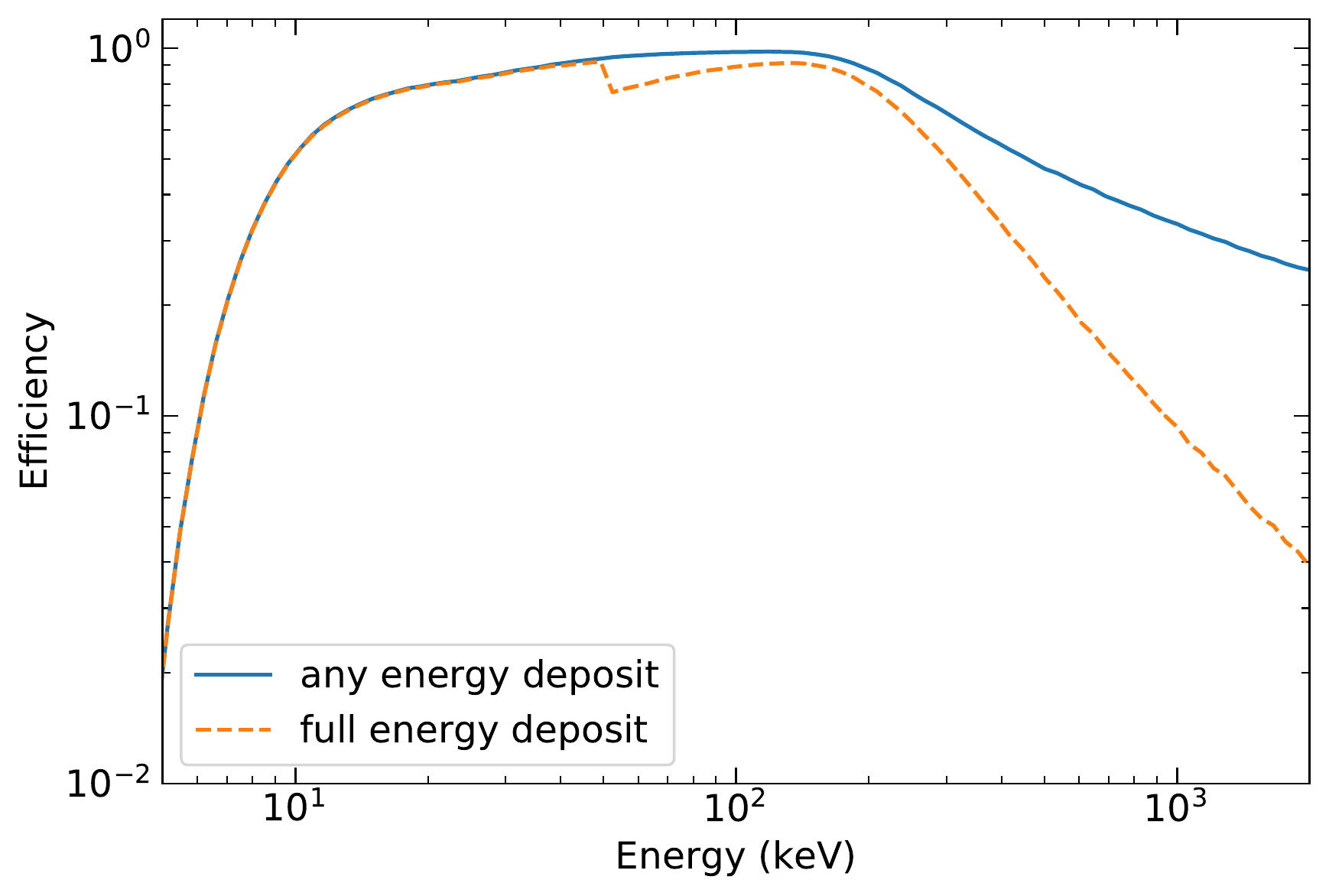}
\caption{Detection efficiency versus energy of the GRID detector. The solid and dashed lines indicate the efficiency for any and full energy deposit, respectively.}
\label{fig:eff}
\end{figure}

Scintillation crystals coupled with silicon photomultipliers (SiPMs) \cite{Ulyanov2016,Murphy2018} are used as the gamma-ray detectors for GRID. The detector including the frontend and backend electronics and the mechanical structure can be packed into a geometry of 9.4~cm $\times$ 9.4~cm $\times$ 5.0~cm, occupying half of the standard unit of the CubeSat (Figure~\ref{fig:detector}). 

The crystal scintillator, Ce-doped Gd$_3$(Al,Ga)$_5$O$_{12}$ (GAGG), is used as the gamma-ray sensor. GAGG has the advantages of high light yield ($\sim$30-70 ph/keV) and high effective atomic number \cite{Iwanowska2013,Yoneyama2018}.  Notably, it shows good mechanical characteristics and is non-hygroscopic. Consequently, the machining and assembly can be less complicated. The detector plane is divided into four identical units, each with an independent setup including the GAGG crystal, SiPM array, and electronics. The unit GAGG crystal has a surface area of $3.8 \times 3.8$ cm$^2$ and a thickness of 1~cm, constituting a total detection area of around 58~cm$^2$ for a single GRID detector. 

The enhanced specular reflector (ESR) manufactured by 3M is adopted as the reflection layer of the crystal due to its high reflectance ($>98$\%) and thinness (65~$\mu$m polymer). It also serves as the entrance window for $\gamma$-rays. The total detection efficiency of the detector is determined by the quantum efficiency of the scintillator, and the transmission of the window and the thermal coat (multilayer polymer) of the whole satellite. For face-on $\gamma$-rays, the detection efficiency is calculated using Geant4 and shown in Figure~\ref{fig:eff}. The angular response of the detector at different energies are also computed, considering both the detector and surrounding medium of the CubeSat, which will be reported elsewhere. 

The SiPM J-60035 manufactured by SensL\footnote{http://sensl.com/downloads/ds/DS-MicroJseries.pdf} is adopted as the optical sensor to measure the scintillation light \cite{Bloser2013}. It shows a photon detection efficiency curve that has a reasonable match with the emission spectrum of Ce-doped GAGG. The SiPM operates at a bias voltage of 28~V, much lower than what is needed for a photomultiplier tube. Each SiPM chip has a geometry of 6.13~mm $\times$ 6.13~mm, consisting of 22,292 single-photon avalanche diodes with a fractional effective area of 75\%. An array of $4 \times 4$ SiPM chips is used in each unit. The outputs of the 16 SiPMs are connected directly and fed into a current sensitive preamplifier, followed by a shaping amplifier, a peak holder, and an analog-digital converter (ADC). This design simplifies the readout electronics but the dark currents from different SiPM chips are co-added. All of the SiPM chips are mounted on the same printed circuit board (PCB). Silica gel is used for optical coupling. All the electronics is controlled and managed by a microcontroller (MCU) ARM Cortex M0+, which communicates with the payload computer via a serial peripheral interface (SPI). A schematic drawing of the GRID detector is illustrated in Figure~\ref{fig:detector}.

The dead time of the electronics is about 50 $\mu$s for the current design. This is mainly due to a relatively slow ADC and MCU that we are using. In the second design, the dead time can be reduced to 15 $\mu$s, which allows for a maximum detection rate of 800 counts~cm$^{-2}$~s$^{-1}$ (with 50\% of event loss). The pile-up effect is negligible, as the preamplifier will be disabled once there is a trigger. Such a design may simplify the algorithm for dead time correction, but considerable event loss during the peak of intense bursts (e.g., TGFs) may happen.

\subsection{Preliminary test results}

\begin{figure}
\centering
\includegraphics[width=0.4\textwidth]{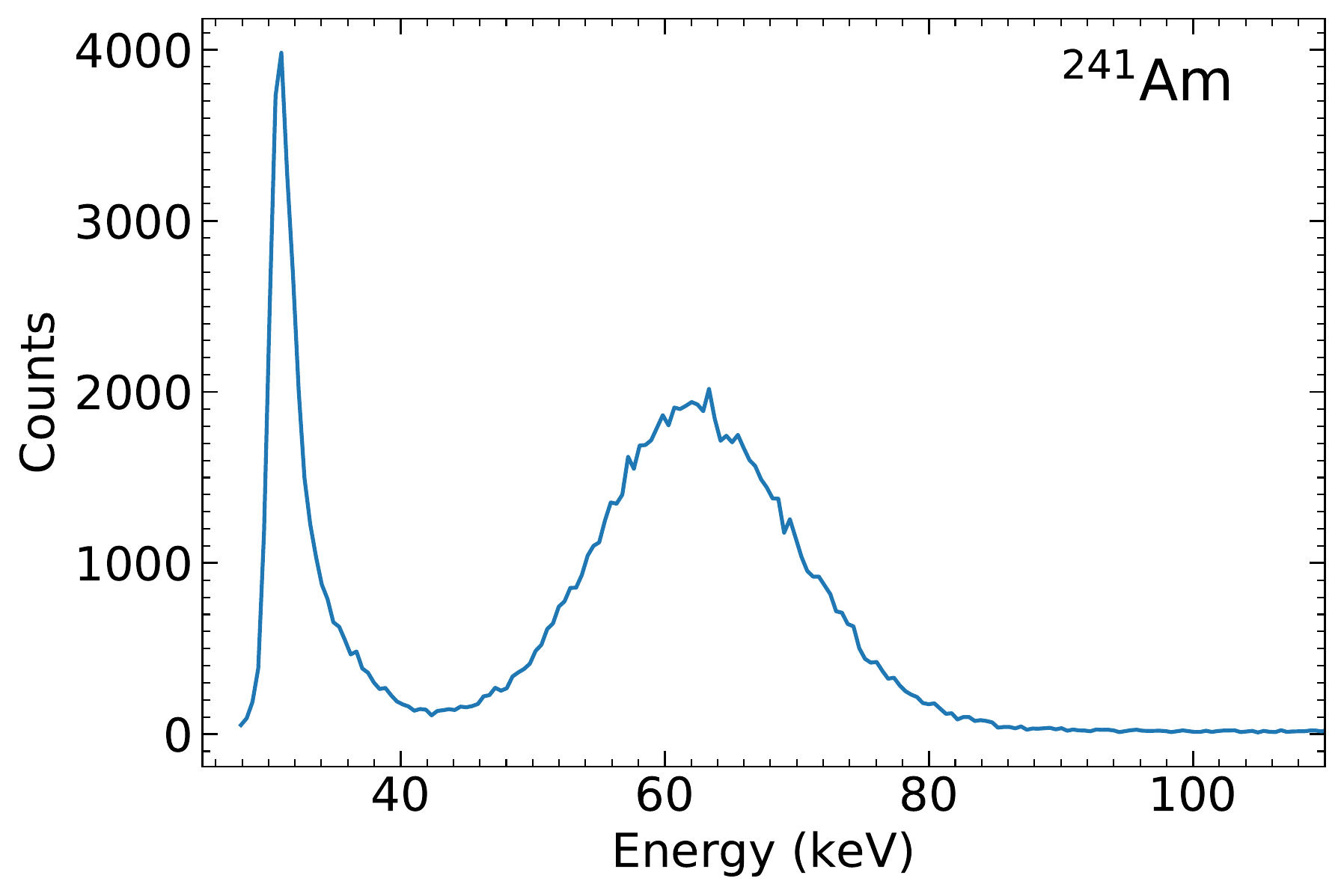}
\includegraphics[width=0.4\textwidth]{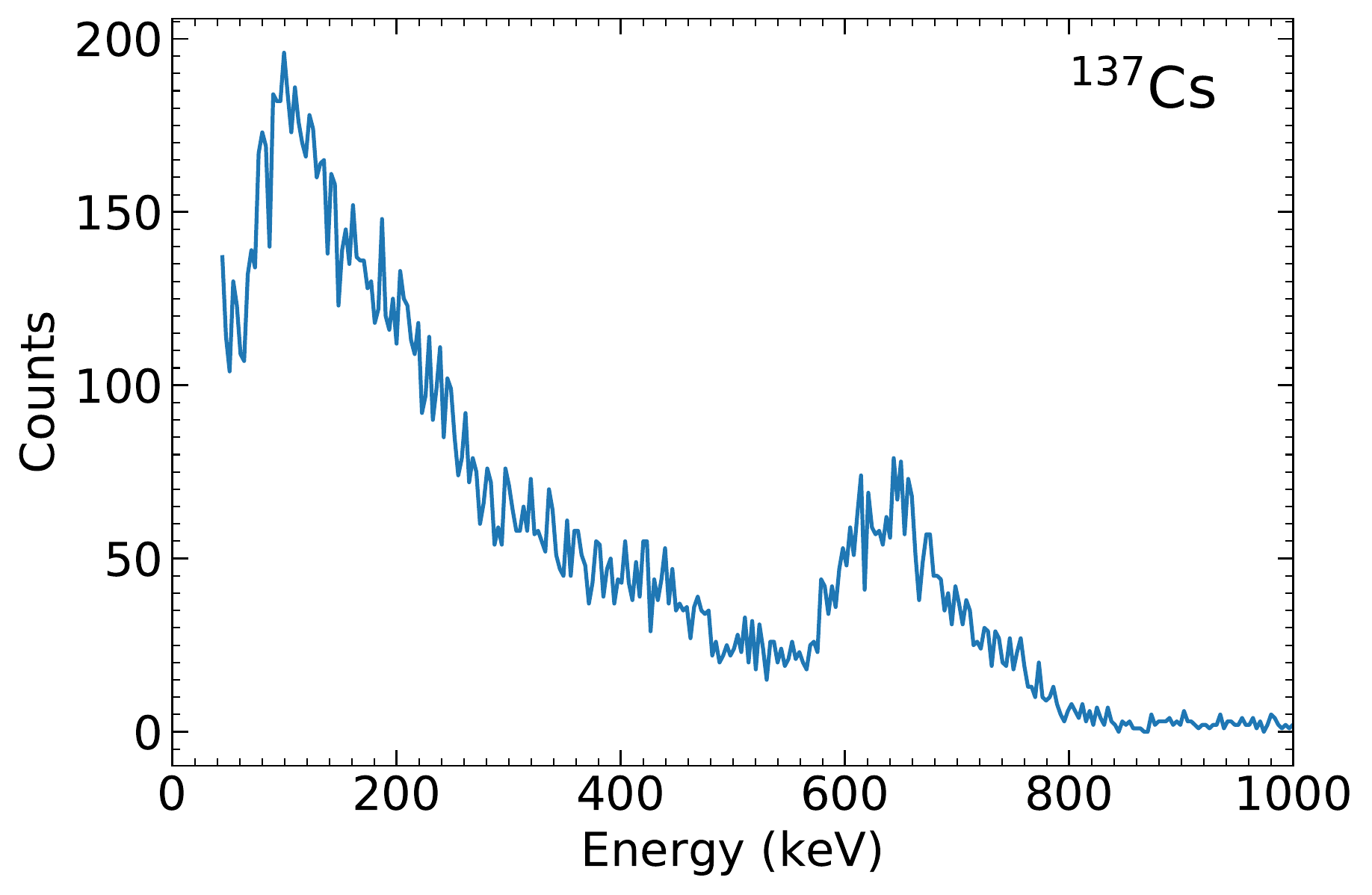}
\includegraphics[width=0.4\textwidth]{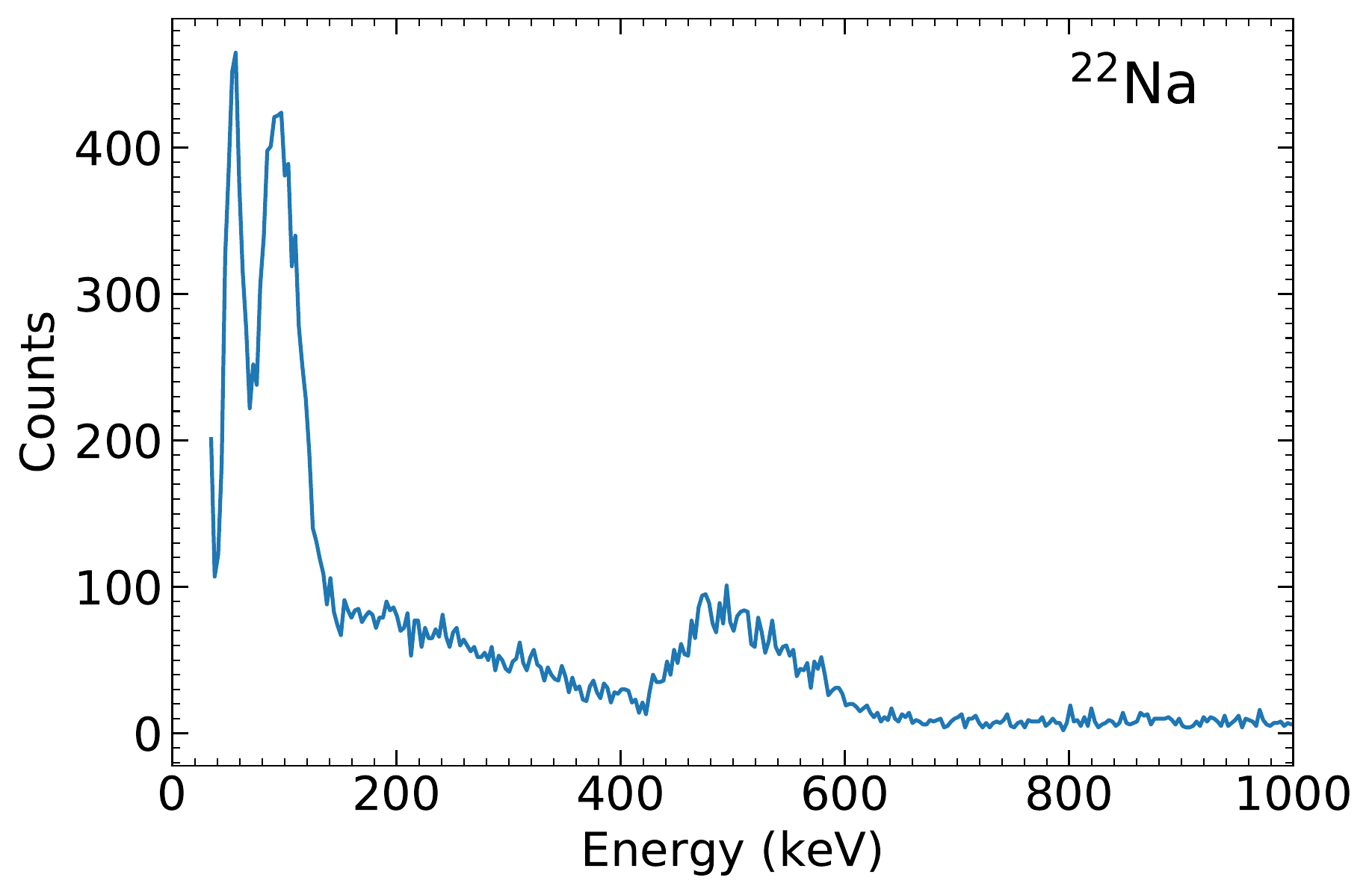}
\includegraphics[width=0.4\textwidth]{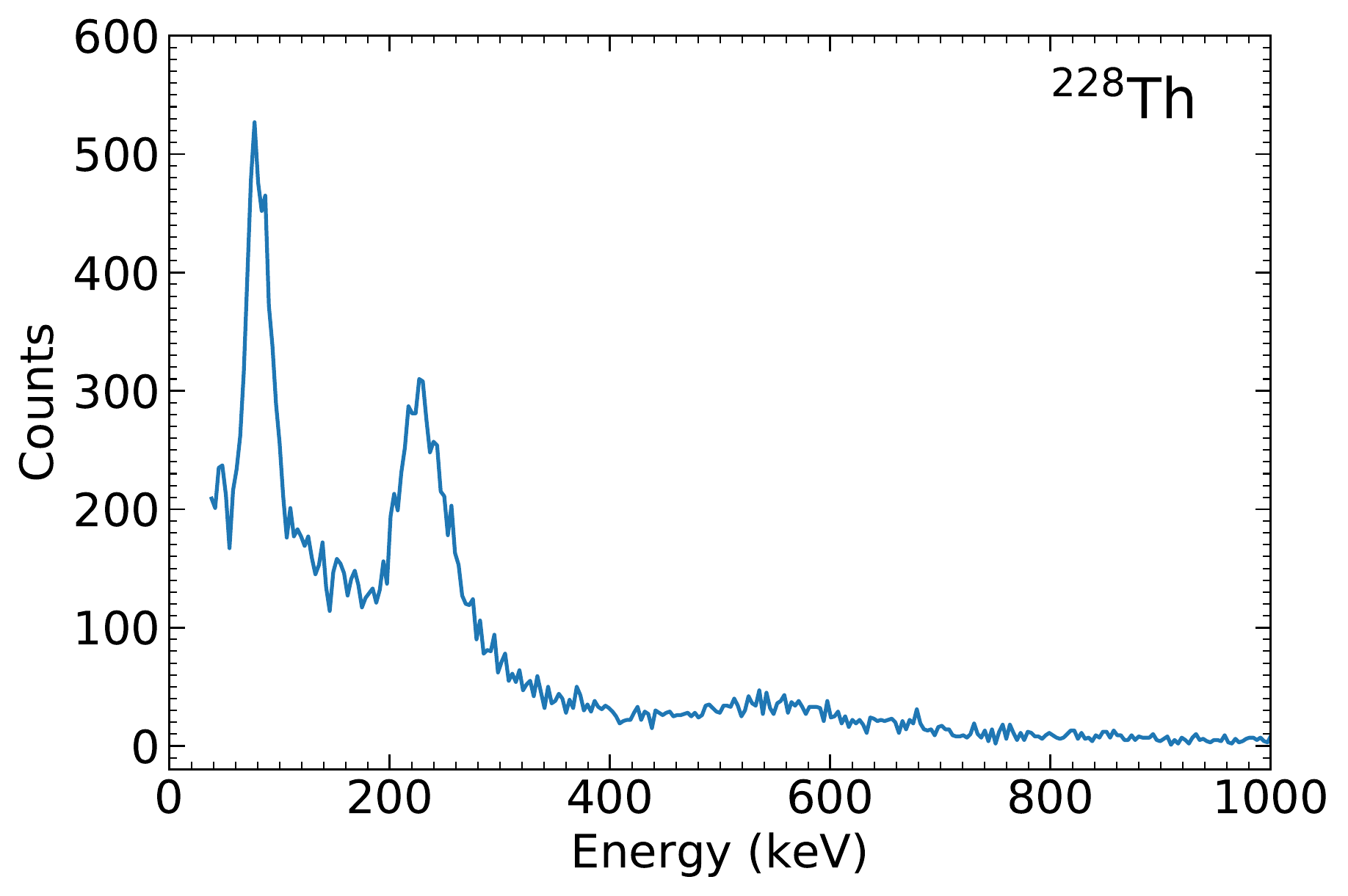}
\caption{Energy spectra measured with radioactive sources. }
\label{fig:spec}
\end{figure}

\begin{figure}
\centering
\includegraphics[width=0.6\textwidth]{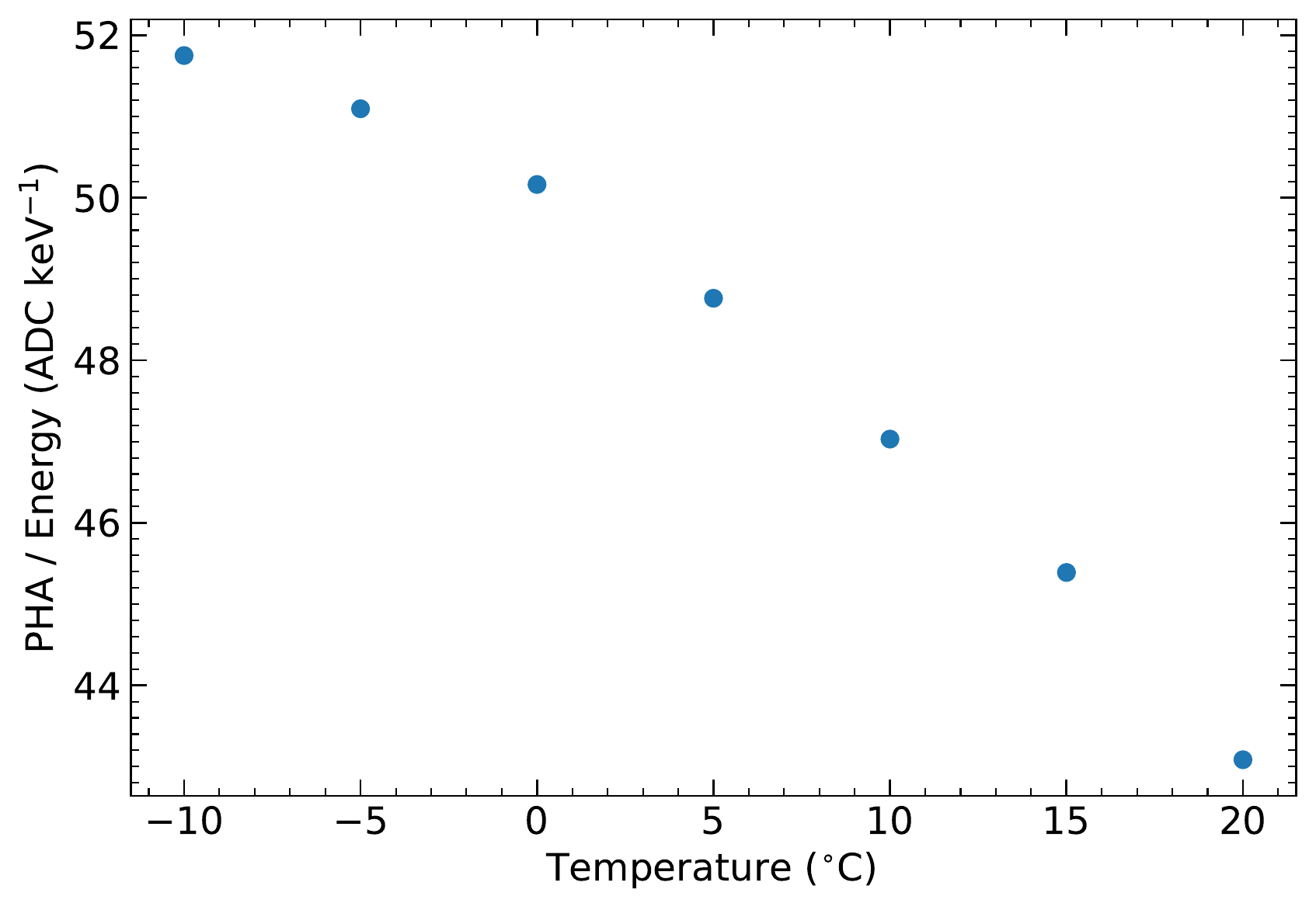}
\caption{Detector gain as a function of temperature. The gain is measured as the ratio of the pulse height amplitude (PHA) to the photon energy, which is the slope of the channel-energy relation.}
\label{fig:cal}
\end{figure}

Energy spectra measured with radioactive sources $^{241}$Am, $^{137}$Cs, $^{22}$Na, and $^{228}$Th are displayed in Figure~\ref{fig:spec}. The energy resolution at 662~keV is measured to be around 20\% (full width at half maximum to the centroid). This is much worse than expected (10\% or better for the material that we are using), mainly due to non-uniformity of light collection in the current design, which will be improved in the future. As the gain of the SiPM is a function of temperature, the channel-energy relation will vary with temperature. The temperature in the spacecraft can be stabilized into a range of 10~$^\circ$C, considering different solar angles and power status (on/off). The temperature of the SiPM will be monitored and used for gain correction. The gain of the detector, i.e., the slope of the channel-energy relation, measured at different temperatures is shown in Figure~\ref{fig:cal}. The measured gain-temperature dependence is lower than those reported in the literature, where a different SiPM is used \cite{Tur2010,Grodzicka-Kobylka2017}, thanks to a smaller temperature coefficient of the SensL J-Series. 

\subsection{Satellite and orbits}

The idea of GRID is to build an extensible collaboration under certain agreements. After two or three launches, we will finalize a standard design for the $\gamma$-ray detector, which can be mounted on future CubeSats with no or little modification to the interface. In such a way, identical $\gamma$-ray detectors can be deployed into various orbits to form a monitoring network. The first GRID payload is mounted on a 6U CubeSat developed by Spacety Co.\ Ltd., a commercial satellite company in China, and was launched into a Sun-synchronous orbit on October 29, 2018. Spacety is responsible for the operation, telemetry, and data transfer.

In most cases, we are not able to choose the orbit for GRID because CubeSats are usually launched as piggyback satellites. In China, dozens of satellites are launched every year and the majority of them go into polar orbits. The polar orbits are not suitable for high energy astrophysical observations, as the trapped particle flux is high at both the polar region and the SAA region, so is the particle-induced background. This results in a shorter observational time and probably a higher background, but could be overcome by utilizing a large number of the GRID satellites in the future.

The trigger algorithm for individual detectors is similar to that for previous GRB monitors, such as {\it Fermi/}GBM. The general idea is that, when the count rate exceeds a certain threshold above the background level, a trigger is announced and lightcurves in several energy bands will be produced as part of the triggered data package. Weaker triggers (e.g., by fainter GRBs) can be identified by cross-correlating the data from different CubeSats in the GRID network. The triggered data can be transferred onto the ground station using the UHF channel in a relatively shorter time. The full data in the event format including the background can be transferred onto the ground in the S or X band.

As a GRB monitor, one of the significant disadvantages of GRID appears to be the non-real time data transfer. It is difficult for GRID to use relay satellites for data transfer because of the limited power for a CubeSat. Ground stations seem to be the only option. The orbital period is roughly 90 minutes for low Earth orbits. If there are sufficient ground stations that can communicate with the majority of the satellites in every orbit, downloading the data with a delay of about an hour is possible. With this amount of time delay, it is still possible to catch the peak of the kilonova after the merger of binary neutron stars. One other possible means for a quick data transfer is to broadcast the small data package after the trigger using the UHF channel, so that radio amateurs all over the world may participate in the program and help collect the data from different satellites. Such an approach needs dedicated efforts on software development and management.

\section{Scientific capability and objectives}
\label{sec:sci}

The primary science drive for GRID is to detect gamma-rays associated with GW events. We ran simulations to estimate the positioning accuracy for these events. We assume 10 satellites evenly distributed in low Earth orbits at an altitude of 600~km and all the detectors are pointing toward the anti-Earth direction. We note that, in reality, a significant fraction of the satellites may crowd in the Sun-synchronous orbits. As a result, the spatial distribution cannot be random. In addition, as our detector is usually not the primary payload of the satellite, they will not always be pointing at the anti-Earth direction. In the following, We choose one of the brightest short GRB in the {\it Fermi/}GBM catalog, GRB 120323A, to roughly estimate the capability of GRID. We first fit the lightcurve of GRB 120323A with two fast rise and exponential decay components, and use it as the template for the lightcurve in our simulations. The background is estimated from the burst-free intervals of this GRB. Then, both the burst and background fluxes are scaled to the GRID band (10--2000 keV) and to the GRID detector area (58~cm$^2$) based on the GBM energy spectra. The background estimated in this way is about 9 counts~cm$^{-2}$~s$^{-1}$ from 10 keV to 2 MeV, slightly higher than but on the same order of magnitude of the rate expected from the cosmic X-ray background (CXB) \cite{Gehrels1992}. This background rate suggests a $5\sigma$ fluence sensitivity for a single detector of about 3.8~photons~s$^{-1}$ or $1.8 \times 10^{-6}$~\ergcm, assuming a typical short GRB spectrum \cite{Nava2011} with a duration of 2~s.

% Band spectrum for SGRB, alpha = -0.5, beta = -2.2, Ec = 489.779 (Nava et al. 2018)
% Model Flux    332.45 photons (0.0001587 ergs/cm^2/s) range (10.000 - 2000.0 keV)
% average detection efficiency in 10-2000 keV = 246 counts / 332 photons
% 5-sigma sensitivity (2 s duration) = 5. * np.sqrt(9. * 58. * 2.) / 58. / 246. * 332.
% or 5. * np.sqrt(9. * 58. * 2.) / 58. / 246. * 0.0001587

\begin{figure}
\includegraphics[width=0.6\textwidth]{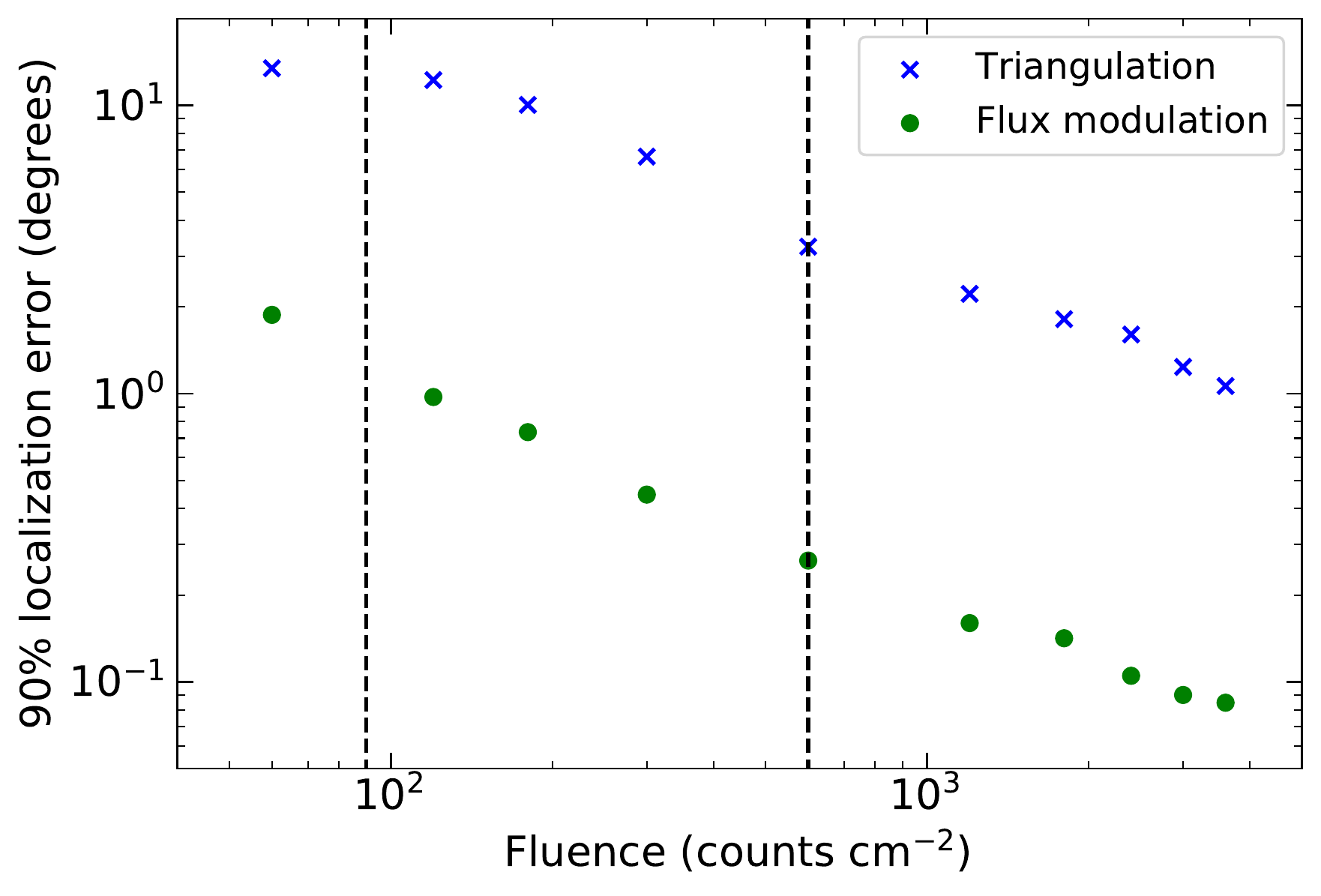}
\caption{Estimated localization accuracy with ten detectors evenly distributed in 600~km low Earth orbits. The dashed line marks the fluence of GRB 170817A ($\sim$90~\ctscm) and 120323A ($\sim$600~\ctscm), respectively. }
\label{fig:error}
\end{figure}

\begin{figure}
\includegraphics[width=0.6\textwidth]{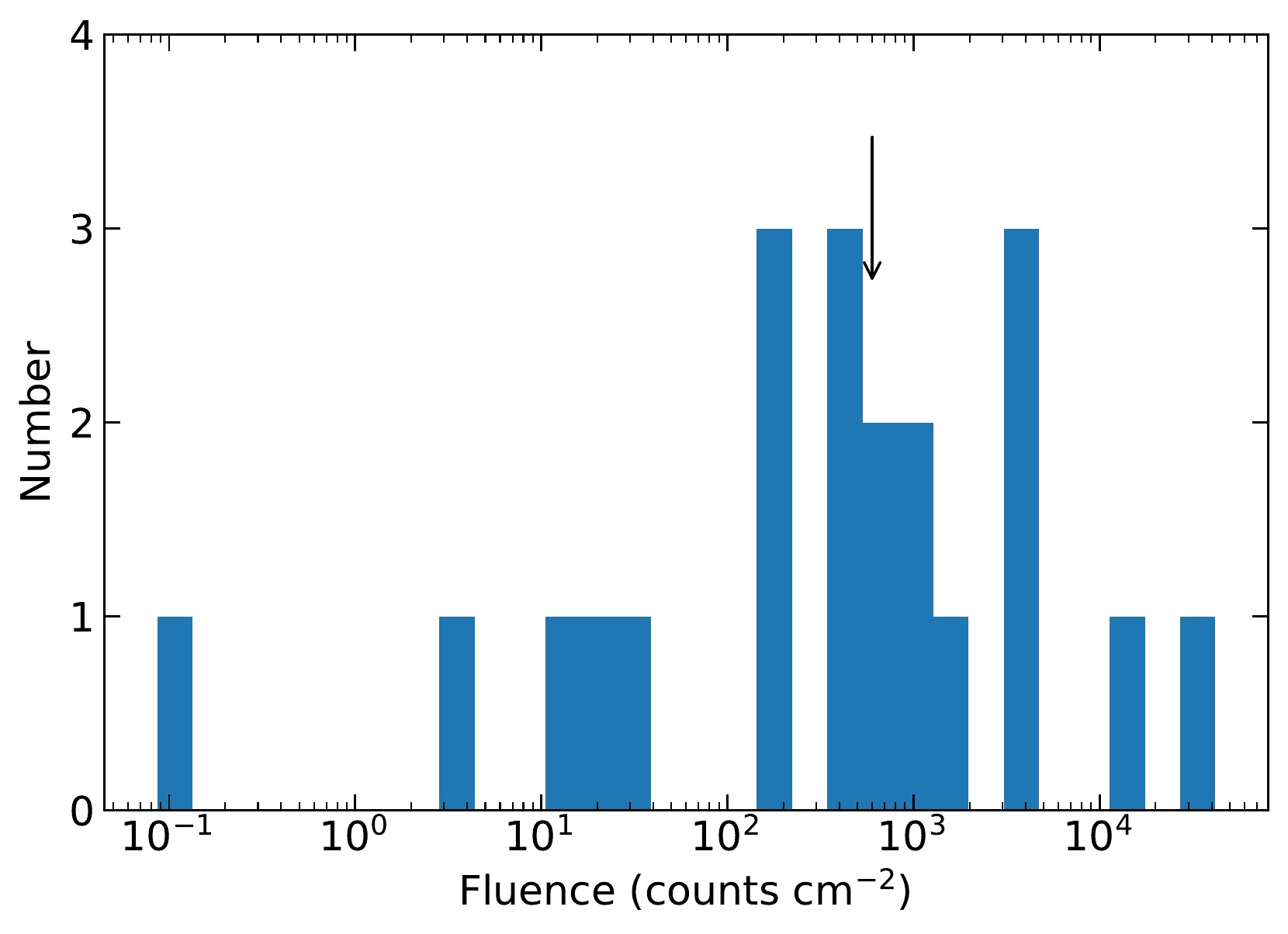}
\caption{Fluence distribution of short GRBs with known redshift if they were at 200 Mpc. The arrow marks the measured fluence of GRB 120323A.}
\label{fig:fluence}
\end{figure}

Given the above assumptions, we simulated mock GRB data in each detector and reconstructed the GRB location using the two techniques. For the triangulation method, the modified cross-correlation function~(MCCF)~\cite{Li2001} is employed to find the time delays with respect to the detector with the highest flux. Compared with the traditional CCF, the MCCF enables a time resolution as high as the instrumental resolution no matter how large the time bin is. For the flux modulation method,  $\chi^2$ minimization is used to find the source coordinate based on the measured flux and flux error in each detector. Given a group of parameters, 100 times of simulations were performed and the 90\% uncertainty of the location distribution was calculated as the positional error. By changing the brightness of the burst, we obtained the localization error as a function of the burst fluence (Figure~\ref{fig:error}). As one can see, for 170817A like GRBs, the localization error is estimated to be about 12$^\circ$, and for 120323A like GRBs, the localization error is about 0.5$^\circ$ to 3.5$^\circ$ depending on the method. Interestingly, if we shift all of the short GRBs with redshift measurements to the distance of 200~Mpc, we find that GRB 120323A has a typical fluence in this sample (Figure~\ref{fig:fluence}). This implies that GRB 120323A may represent a standard GRB with a jet pointing toward us at a distance close to the horizon of the advanced LIGO in the near future. The flux modulation method always produces a better accuracy than the triangulation method. However, as we have assumed ideal conditions in the simulation, the accuracy may have been underestimated, especially for the flux modulation method, which will be affected by scattering on the satellite and the Earth atmosphere. On the other hand, the triangulation method is less sensitive to these effects. 

We note that, roughly speaking, the positioning accuracy scales inversely with the square root of the total number of satellites. It is not a strong function of the detector orientation, being random or anti-Earth, but will be affected by the orbital distribution. The accuracy will decrease,  more with the triangulation method than flux modulation, if all satellites are clustered in a single orbit. This is not difficult to understand because the flux modulation method does not rely on the positions of the satellites.  Also in this extreme case, the accuracy is better if the source is perpendicular to the orbital plane, no matter which method is used, because more satellites can see the burst.  Here, we just present some rough estimates and this topic needs further investigation when the detector performance and the orbital configuration are better known \cite{Ohno2018}.

\subsection{Estimated event rate}

In this subsection, we calculate the event rate for GRBs that can be jointly detected by GRID and ground-based GW detectors. For the advanced LIGO, the horizon in 2020 for neutron star mergers is estimated to be about 200~Mpc~\cite{Abbott2016}, which is adopted as the upper limit for the distance. In the following, we discuss the expected event rate in two cases, for either the standard, on-axis GRBs or 170817A-like GRBs.

(1) \textit{Standard on-axis GRBs}. In this case, the GRB jet will point toward us, the event rate can be estimated from their local density. The redshift dependent luminosity function for short GRBs can be modeled by adding a time delay (since binary formation to merger) distribution onto the star formation history at different redshifts \cite{Sun2015,Wanderman2015}. The number density of short GRBs is a function of both the redshift and luminosity, while the observed number is also subject to the sensitivity of the detector. For standard, on-axis short GRBs within 200~Mpc, it is reasonable to assume that the least luminous burst can trigger our detectors. The isotropic luminosity function of the existing short GRBs (except GRB 170817A; likely an off-axis example) cuts at about $7 \times 10^{49}$~\ergs. The local short GRB number density can be derived by integrating the luminosity function, which is found to be a power-law with an index of $-1.7$ if a Gaussian time delay is assumed \cite{Sun2015}, from the threshold luminosity to infinity, and is found to be \cite{Sun2015}
\begin{equation}
\rho_0 (L_{\rm iso} > 7 \times 10^{49} \; {\rm erg\; s}^{-1}) = 4.2_{-1.0}^{+1.3} \;\; {\rm Gpc}^{-3} \; {\rm yr}^{-1}.
\end{equation} 
This can be translated to an event rate of
\begin{equation}
n (< 200 {\rm Mpc}) = 0.14_{-0.03}^{+0.04} \;\; {\rm yr}^{-1}.
\end{equation} 
Different model assumptions may produce different results, but they give consistent results at the same order of magnitude \cite{Sun2015,Wanderman2015}. 

(2) \textit{170817A-like GRBs}. GRB 170817A has an isotropic luminosity of only $1.7 \times 10^{47}$~\ergs. It is still unknown whether the $\gamma$-ray emission is intrinsically faint or most of the power is beamed away from our line of sight. In any case, the detection of GRB 170817A offers a direct estimate of the event rate by taking into account the GBM operating time, field of view, and sensitivity. Such calculation gives a local density of short GRBs with an isotropic luminosity as low as that of this event \cite{Zhang2018},
\begin{equation}
\rho_0 (L_{\rm iso} > 1.7 \times 10^{47} \; {\rm erg\; s}^{-1}) = 190_{-160}^{+440} \;\; {\rm Gpc}^{-3} \; {\rm yr}^{-1},
\end{equation} 
or an event rate
\begin{equation}
n (< 200 {\rm Mpc}) = 6_{-5}^{+15} \; {\rm yr}^{-1}.
\end{equation} 
We note that an integration in the short GRB luminosity function (in case 1) down to this low threshold leads to a consistent result, which suggests that this low luminosity event lies right on the extension of the short GRB luminosity function. 

An alternative way to estimate the event rate is based on the neutron star merge rate. The detection of GW170817 in the two runs O1 and O2 gives a neutron star merger rate \cite{Abbott2017}
\begin{equation}
\rho_{0, {\rm NS-NS}} = 1540_{-1220}^{+3200} \;\; {\rm Gpc}^{-3} \; {\rm yr}^{-1}.
\end{equation} 
The off-axis angle for GRB 170817A is argued to be about 30$^\circ$, consistent with the jet opening angle for short GRBs \cite{Fong2015}. This corresponds to a fractional sky coverage of 0.07 and a detection rate
\begin{equation}
n (< 200 {\rm Mpc}) = 4_{-3}^{+8} \; {\rm yr}^{-1}.
\end{equation} 
Such a value is consistent with the above estimate and suggests that GRID may observe a few short GRBs per year associated with neutron star mergers. 

\subsection{Other GRB science}

For traditional GRB studies, GRID can be treated as similar to GBM but with individual detectors distributed around the low Earth orbits. Compared with GBM, the whole GRID network with 10-20 CubeSats will have a similar energy range, spectral resolution, and sensitivity. GRID is estimated to catch around 500 GRBs every year, which offers a better unbiased larger sample in GRB research. In addition, thanks to its all-sky coverage, ultra-long GRBs and ``prototype'' events such as GRB 160625B can be well-observed. 

\subsection{Magnetars}

Magnetars are neutron stars with extremely strong magnetic fields \cite{Duncan1992}. By now, only 29 sources (including 6 candidates) have been discovered \cite{Olausen2014}. Therefore, it is needed to enlarge the sample in order to dig deeper in the physics of these extreme objects. Catching bursts in soft gamma rays from magnetars is a very efficient way to find new objects. GBM has detected 446 magnetar-like bursts from 2008 to 2015, among which 427 were found to be produced by 8 known sources, while the origin of the other 19 bursts could not be identified due to coarse localizations \cite{Collazzi2015}. Based on their peak fluxes and durations, the location of these unidentified bursts can be reconstructed to an accuracy of about 1--5 degrees with GRID. Identification of such bursts requires follow-up observations with wide field X-ray telescopes such as the Einstein Probe~\cite{Yuan2018}. Similar to GBM, we expect to detect bursts from $\sim$3 magnetars per year with GRID. 

\subsection{TGFs}

TGFs are produced in the Earth atmosphere, featured with microseconds to milliseconds duration, single or multiple pulses, and very hard spectrum (up to 100 MeV or higher) \cite{Briggs2010,Tavani2011}. The association of TGFs with thunderstorm regions \cite{Fishman1994} and lightning discharge \cite{Inan1996,Cohen2006} has been confirmed. Along with the production of TGFs, it is suggested that some fraction of the secondary electrons and positrons may escape the atmosphere and be captured by the geomagnetic field. Those electrons and positrons will then travel along the geomagnetic field lines, forming the TEBs \cite{Dwyer2003}. Due to the atmosphere absorption, TGFs could only be detected within $\sim$800~km from the production site, while TEBs can be observed in several thousand kilometers away \cite{Briggs2011}. Detecting TGF with one satellite and TEBs with others at almost the same time will be a direct test of the TGF and TEB emission and propagation model. A rough estimate indicates that GRID can detect about $\sim$$10^3$ TGFs per year. Compared with a single satellite, a large number of GRID satellites with high inclination orbits will increase the probability of detecting TGFs in full latitude range and associated TGFs and TEBs.

\subsection{A student project}

The GRID concept was first proposed in October of 2016 by a group of undergraduate students, inspired by discussions with several professors. Then, a student team dedicated to this project was organized, consisting of mainly undergraduate students and a few graduate students. Due to the technical readiness of $\gamma$-ray detectors, this project is indeed suitable for training students, not only how to develop instruments, but also how to lead and organize an inter-discipline collaboration. So far, GRID has been led by students all the way from science justifications and simulations, hardware design and testing, data analysis, and paper publication. In modern experimental physics and astronomy, large science projects with collaborations from many institutes/countries and disciplines have become more and more popular. The gravitational wave detection itself is a remarkable example. How to attract students in different majors into fields of basic research, how to train them to think of and solve problems professionally, and eventually, how they can grow to be leaders for future large science projects, are questions of great interest.

With the GRID project, we will explore a new way for student education and training. We offer the students, especially undergraduate students, freedom to choose the research topic of their interest  and conduct the research independently and professionally. Unlike the traditional way that every graduate student is associated with a supervisor, graduate students in GRID may be treated as independent researchers. Of course, guidance is still needed, but the whole supervisor team instead of individuals can act as the advisor. 

GRID is a cost-effective project and is currently supported by university funds. At the timing of writing, students from more than 20 institutes have participated in the GRID collaboration; the supervisors are mainly from but not limited to their host institutes.

\section{Summary and future perspectives}

To summarize, GRID is a student project with a dedicated and straightforward scientific goal --- to detect and locate GRBs produced by neutron star mergers jointly with ground-based GW detectors in the local universe. The purpose of the first launch is to test the hardware design, trigger algorithm, and data transfer. In the meanwhile, we have constructed a Geant4 package to simulate the detector response, as a function of photon energy and incident angle, and to model the in-orbit background. We will also take into account the atmospheric scattering by including an Earth atmosphere mass model. The simulation results and comparison with laboratory tests will be elaborated in a separate paper. In the near future, with one or two more launches and other laboratory tests, we will deliver and share a package inside the collaboration that includes the following items.

\renewcommand{\labelitemi}{$\bullet$}
\begin{itemize}

 \item Detailed payload design. This includes the mechanical drawings, electronic design and associated firmware, a list of components with suggested suppliers, and technology and requirements for detector assembly. Based on such information, any member of the GRID collaboration can make an identical copy of the payload. 
 
 \item A guideline for experiments, calibrations and software developments. In practice, the calibrations can be done in some member institutes where the facilities are ready. 
 
 \item The Geant4 simulation package. As the structure of the satellite will change, a team will be responsible for building and running a simulation package for each satellite to create response files for data analysis.
 
 \end{itemize}

A standard design that is compatible with most of the commercial CubeSats will significantly lower the cost. If every member institute can contribute one or a few detectors in the following years, the GRID network can be quickly formed and turned into operation mode. Plus, if most of the member institutes can host a UHF ground station, the time delay from a trigger to the reception of the small data package could be reduced to within one orbital period. In general, the GRID collaboration resembles a decentralized architecture and the members in the collaboration have equal status.
  
\begin{acknowledgements} 
We thank the referee for useful comments. HF acknowledges funding support from the National Natural Science Foundation of China under the grant Nos.\ 11633003 \& 11821303, and the National Key R\&D Program of China (grant Nos.\ 2018YFA0404502 and 2016YFA040080X). BBZ acknowledges support from National Thousand Young Talents program of China and National Key Research and Development Program of China (2018YFA0404204) and The National Natural Science Foundation of China (grant No. 11833003). This work is supported by Tsinghua University Initiative Scientific Research Program.
\end{acknowledgements}

%\bibliographystyle{spmpsci} % mathematics and physical sciences
%\bibliography{grid,eirsat}

\end{document}